# Measurement of the Dzyaloshinskii-Moriya Interaction in Mn$_4$N Films that Host Skyrmions

Wei Zhou [1], Chung Ting Ma [1,*] and S. Joseph Poon [1,2]

1. University of Virginia, Department of Physics, Charlottesville, Virginia, 22904, USA
2. University of Virginia, Department of Materials Science and Engineering, Charlottesville, Virginia, 22904, USA
*ctm7sf@virginia.edu

## ABSTRACT

Mn$_4$N thin film is one of the potential magnetic mediums for spintronic devices due to its ferrimagnetism with low magnetization, large perpendicular magnetic anisotropy (PMA), thermal stability, and large domain wall velocity. Recent experiment confirmed the existence of tunable magnetic skyrmions in MgO/Mn$_4$N/Cu$_x$Pt$_{1-x}$(x=0,0.5,0.9,0.95), and density functional theory (DFT) calculation provided a large theoretical value of the interfacial Dzyaloshinskii-Moriya Interaction (iDMI) of Mn$_4$N/Pt, which is consistent with the predicted chemical trend of DMI in transition metal/Pt films. So far, measured DMI has not been reported in Mn$_4$N which is needed in order to support the predicted large DMI value. This paper reports the average DMI of MgO/Mn$_4$N(17nm)/Cu$_x$Pt$_{1-x}$(3nm), extracted from the anomalous Hall effect with various tilted angles, based on magnetic droplet theory with DMI effects. The DMI decreases from 0.267 mJ/m$^2$ to 0.011 mJ/m$^2$ with non-linear tendencies as Cu concentration in Cu$_x$Pt$_{1-x}$ capping layer increases from 0 to 1, demonstrating the control of DMI through Cu$_x$Pt$_{1-x}$ capping layer. Furthermore, a solid solution model is developed, based on X-ray photoelectron spectroscopy (XPS) compositional depth profile, to analyze the possible effects on DMI from the mixing layers at the surface of Mn$_4$N. After taking into account the mixing layers, the large DMI in Mn$_4$N film with Pt capping is consistent with the predicted DMI.

## Introduction

As information technologies keep developing, the demand for faster processing and high-density data storage is increasing [1,2]. Spintronics device, which utilizes and manipulates the spin degrees of freedom in materials, is a promising candidate for next-generation energy-efficient electronic devices with high-speed operation and ample data storage.[1-5]. In a spintronic device, the magnetic medium is the critical component that determines the device's performance [2-5]. Ferrimagnetic systems with two antiparallelly coupled spin sublattices have drawn increasing attention for two reasons. The first reason is that ferrimagnetic materials have faster switching processes than ferromagnetic materials [6,7]. The other reason is the high-speed current-induced magnetic domain-wall motion in ferrimagnets. [8-10]

Among numerous ferrimagnetic materials, anti-perovskite $Mn_4N$ thin films with a Curie temperature of 710K have attracted more investigations recently [11,12]. Figure.1 shows the schematic diagram of the $Mn_4N$ crystal structure with spins. The spins of Mn I atoms (3.47 $\mu_B$), which sit at the corners, are ferromagnetically coupled with the spins of Mn IIa atoms (0.75 $\mu_B$), which sit at the face-center of the top and bottom surfaces in the unit cell. The spins of Mn I and Mn IIb atoms ($-2.36$ $\mu_B$), which sit at the face-center of side surfaces in the unit cell, are anti-ferromagnetically coupled [13]. It has been reported that the epitaxial $Mn_4N$ thin films grown on various substrates, such as MgO(001), STO(001), $LaAlO_3$(001), and LSTO(001)[10,13-19], exhibit perpendicular magnetic anisotropy (PMA), which is essential for some spintronic devices. The magnetization of $Mn_4N$ thin films is tunable by doping Ni ($Mn_{4-x}Ni_xN$) or Co ($Mn_{4-x}Co_xN$), and magnetic compensation (zero net magnetization) can be achieved with suitable Ni or Co composition [20-22]. More importantly, compared to ferrimagnetic rare-earth transition metal amorphous thin films, the $Mn_4N$ thin films have better thermal stability because $Mn_4N$ films are deposited at 400-450 °C, and no structural transitions or loss of PMA has been reported after annealing and cooling processes. Furthermore, experiments have reported high domain wall velocity(~1km/s) [10,23], high spin polarization (0.8) [22], and magnetic skyrmions in Mn$_4$N [23], which indicates that the Mn$_4$N thin film is a potential material for spintronic devices such as racetrack memory and skyrmion-based magnetic tunnel junctions [24,25]. Besides, the skyrmions' diameter in MgO(001)/Mn$_4$N/Cu$_x$Pt$_{1-x}$ can be tuned by changing the composition of the capping layer, which would vary the interfacial Dzyaloshinskii–Moriya interaction (iDMI) [23].

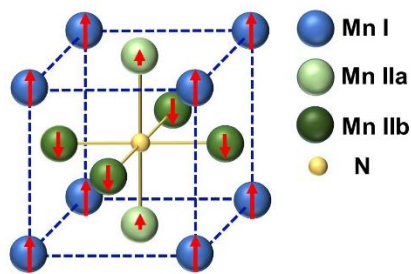

**Figure 1.** Schematic diagram of the Mn$_4$N crystal structure.

The iDMI is an antisymmetric exchange interaction that favors the noncolinear alignments of neighboring spins [26,27]. It arises from the spin-orbit coupling at magnetic layer interfaces with broken inversion symmetry. The iDMI, which has attracted great interest in recent years, is one of the crucial interactions to develop new promising spintronic applications [27]. For example, iDMI is vital in stabilizing topologically non-trivial chiral magnetic textures, such as Néel magnetic skyrmions and chiral domain walls [3,4,26], which are candidates to serve as building blocks in data storage. In addition, the interplay of DMI and spin-orbit torque (SOT) provides a fast and power-saving method of field-free current-induced switching of perpendicular magnetization, which is critical in low-energy and high-speed calculations [28,29].

It has been predicted by first-principles calculations that at the interface of 3d transition metals (TM) (V, Cr, Mn, Fe, Co, Ni) and 5d heavy metal (HM) (W, Re, Os, Ir, Pt), the 3d orbit occupation of TM serves an important role of the DMI strength. The Mn element has the largest DMI due to its half-filled 3d-band [30]. While many DMI measurements have applied to often studied Co and Co-based films with HM interface [31-39], no measured DMI of Mn or Mn-based magnetic films has been reported to support the systematic trend of increasing iDMI. Besides, density functional theory (DFT) has predicted a large iDMI at the $Mn_4N$/Pt interface where DMI is only considered between the Mn nearest neighbors (Mn I and Mn IIa atoms) at the $Mn_4N$/Pt interface.[23] Thus, experimental DMI in $Mn_4N$ is necessary to confirm the calculations and for comprehensively exploring the material for applications. Kim et al. developed a method, based on magnetic droplet theory with the DMI effect, to extract the average DMI from in-plane field dependence of out-of-plane nucleation field for a reversed domain. This method can be completed with a magnetoresistance measurement setup or Magneto-optical Kerr-effect (MOKE) microscope [40]. Here, to confirm the calculated large $Mn_4N$ DMI and comprehensively explore the material for spintronic applications, theDMI of MgO/$Mn_4N$(17nm)/$Cu_xPt_{1-x}$ (x=0, 0.5, 0.9, 1) by extracting the effective field of DMI from angular dependence anomalous Hall effect. The average DMI of the MgO/$Mn_4N$/$Cu_xPt_{1-x}$ decreases non-linearly from 0.267 mJ/$m^2$ to 0.011 mJ/$m^2$ as Cu concentration increases from 0 to 1. Furthermore, MgO/$Mn_4N$(17nm)/Pt has a larger interfacial DMI constant (Ds) than MgO/Co(0.5-1.2nm)/Pt film[31-39], where Ds is the product of average DMI and the thickness of the magnetic layer. The larger Ds of MgO/$Mn_4N$/Pt is consistent with the DMI trend calculated by A. Belabbes *et al.*[30]. Moreover, the possible effect on the DMI of the mixing layer at the surface of $Mn_4N$ is discussed. The multilayers are divided into tens of sublayers using compositional gradient. The composition of each sublayer is estimated from X-ray photoelectron spectroscopy (XPS). A solid solution model is used to calculate the average DMI of the MgO/Mn$_4$N(17nm)/Cu$_x$Pt$_{1-x}$, which incorporates the effect from the mixing layers at the surfaces of Mn$_4$N layers.

## Methods

17 nm thick Mn$_4$N thin films were deposited on the MgO(001) 5 × 5 × 0.5 mm substrate by reactive radio frequency (rf) sputtering at 450 °C. The MgO substrates were wet-cleaned and heat-treated ex-situ. The base pressure was 5x10$^{-8}$ Torr, and the deposition pressure was 1x10$^{-3}$ Torr. The flow rates ratio of Ar and N$_2$ gases was maintained at a flow rate ratio of 93:7. 3 nm thick capping layers of Cu$_x$Pt$_{1-x}$ (where x = 1, 0.5, 0.1, 0) were deposited on the Mn$_4$N layer at room temperature by co-sputtering Pt and Cu targets to tune the DMI. Then, a 3 nm thick Pt layer is deposited on top to prevent oxidation. The structure of the films is shown in Figure.2(a). Details of the deposition process and cleaning of MgO substrates were reported in a previous work [18]. The composition of the capping layers was calibrated with 10 nm thick Cu$_x$Pt$_{1-x}$ films on MgO(100) substrates using PHI VersaProbe III X-ray photoelectron spectroscopy(XPS). The out-of-plane and in-plane magnetic hysteresis loops of each sample were measured at 300 K by a Quantum Design VersaLab III vibrating sample magnetometer (VSM). The Mn$_4$N films were patterned into a 5-μm-wide Hall cross-structure by photolithograph and Ar ion milling technique. A 100 nm thick Pt layer was deposited on the patterned samples as contact pads for anomalous Hall effect (AHE) measurements.

We measured the average DMI using the method proposed by Kim et al., which is based on the magnetic droplet nucleation model [40]. The schematic of the measurement setup is shown in Figure.2(b). Angular-dependent coercivity field H$_c$ of the Mn$_4$N Hall cross-structure was measured. By definition, the perpendicular component (H$_z$) and in-plane component (H$_x$) of H$_c$ are given by H$_z$ ≡ H$_c$ cosθ and H$_x$ ≡ H$_c$ sinθ, where θ is the angle between the external magnetic field (H) and the sample's normal. θ varied from 0° to 65° in this experiment. The external magnetic field was swept within ±2 T at each angle to observe the coercivity field $H_c$. Figure.2(d) shows normalized anomalous Hall effect loops of MgO/Mn$_4$N/Pt with different tilted angles θ. With DMI, there is a threshold point in $H_z^{1/2}$ vs. $H_x$ curve, where $H_z^{1/2}$ begins to decrease linearly with increasing $H_x$. This threshold point corresponds to the effective magnetic field induced by DMI($H_{DMI}$).

XPS measurements were performed to obtain the compositional depth profile using the PHI VersaProbe III XPS instrument. XPS data were collected after sputtering off a few layers from the surface. Each sputtering lasted 15 s, and the total sputtering time was 10 min. The analysis method followed the method in ref[41].

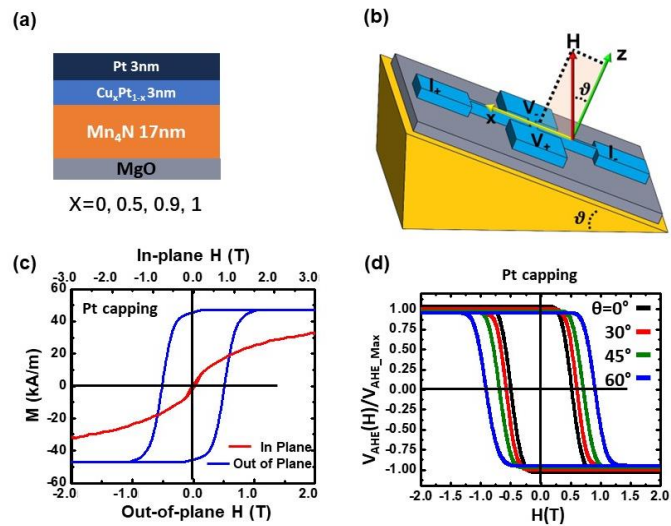

**Figure 2.** (a) Structure of the samples MgO(100)/Mn$_4$N(17nm)/Cu$_x$Pt$_{1-x}$(3nm)/ Pt(3nm). (b) Schematic setup of the DMI measurement. The samples were put on a tilted holder while the external magnetic field (H) was applied in vertical direction. The out-of-plane component $H_z$ and the in-plane component $H_x$ of the coercivity field $H_c$ were calculated with $H_z \equiv H_c \cos\theta$ and $H_x \equiv H_c \sin\theta$. The range of tilted angle θ was 0-65°. (c) M(H) loops of MgO/Mn$_4$N/Pt with out-of-plane external magnetic field and in-plane external magnetic field measured by VSM.

(d) Normalized anomalous Hall voltage ($V_{AHE}$) loops of MgO/Mn$_4$N/Pt with different tilted angles ($\theta$=0° (black), 30° (red), 45° (green), 60° (blue)).

## Results and Discussion

*DMI measurement*

Figure.3(a)-(d) shows the $H_z^{1/2}$ vs. $H_x$ curves of capping layers with different Cu composition (x) (x = 0, 0.5, 0.9, 1). The lines in the graphs are the fitted lines. The turning points correspond to the $H_{DMI}$ of each capping layer. As the Cu concentration of the capping layer(x) increases, the $H_z^{1/2}$ begins to decrease linearly at a lower $H_z^{1/2}$. This means that $H_{DMI}$ decreases and the average DMI decreases. This trend matches the previously reported tunable size of skyrmions in Mn$_4$N with Cu$_x$Pt$_{1-x}$ capping, where the size of the magnetic skyrmion in MgO/Mn$_4$N/Cu$_x$Pt$_{1-x}$ decreases as the Cu concentration increases [24]. This agrees with the intuition that smaller DMI produces smaller skyrmions [24]. The DMIs of different capping layers are calculated based on the following equation:

$$DMI = \sqrt{A/K_{eff}} M_s H_{DMI} \tag{1}$$

where A is the exchange stiffness, $K_{eff}$ is the effective perpendicular magnetic anisotropy energy, and $M_s$ is the saturation magnetization. For Mn$_4$N, A is 18 pJ/m[10], $M_s$ and $K_{eff}$ are 47 kA/m and 1.05×10$^5$ J/m$^3$, respectively, based on the VSM measurement of MgO/Mn$_4$N/Pt. The measured DMIs of different capping layers are plotted in Figure 3 (e). The errors in the measured DMI mostly come from the uncertainties of the tilted angle and fitting. As Cu concentration rises, the DMI decreases from 0.267±0.065 mJ/m$^2$ to 0.011±0.01 mJ/m$^2$. This decrease in DMI can be explained by two reasons. First, the addition of Cu diluted the concentration of Pt, which reduce the large iDMI at interface of Pt/Mn$_4$N. Second, iDMI at Cu/Mn$_4$N interface is small and has a opposite sign of iDMI at Pt/Mn$_4$N interface. This would further decrease the iDMI as Cu concentration increases. We note that the decrease in DMI is not linear as a function of Cu composition x. When x increases from 0 to 0.5, the DMI decreases from 0.267±0.065 mJ/m$^2$ to 0.224±0.053 mJ/m$^2$. Compared to the large change in the CuPt capping layer composition, the change in DMI is small and the difference is within the measurement error. This indicates that when the Cu concentration is smaller than 0.5, the DMI is almost insensitive to the change in Cu concentration. When x further increases to 0.9, the DMI decreases to 0.115±0.041 mJ/m$^2$, which is about half of the DMI when x is 0.5. When it is pure Cu capping (x = 1), the DMI decreases more to near 0. Since small amount Pt in the capping layer can provide a large DMI, it indicates that the DMI is more sensitive to the Pt concentration than Cu concentration. This can be explained by the larger spin-orbit coupling (SOC) between Pt and Mn than the SOC between Cu and Mn [23,42]. The non-linear composition dependence of DMI has also been reported in Pt/CoGd/W$_x$Pt$_{1-x}$ [43]. Since DMIs in these thin films are interfacial effects originating from the interface, the measured DMIs decrease with thicker magnetic layers. To compare the DMI with other materials, which have different thicknesses, we use the interfacial DMI constant Ds, where Ds is the average DMI multiplied by the magnetic layer thickness($t_m$). The result is shown in Figure.3(f), the Mn$_4$N data point is from our measurement of MgO/Mn$_4$N/Pt, and the Co data point is the average of several reported Ds in Co single layer with different thicknesses sandwiched by MgO and Pt, where the error bar is the standard deviation [31-39]. The Ds of MgO/Mn$_4$N/Pt is about twice the Ds of MgO/Co/Pt, which is consistent with the chemical trend of DMI in transition metals from first-principles calculations [30].

The measured DMI of MgO/Mn4N/Pt (0.267 mJ/m2) is one magnitude smaller than DFT predicted iDMI of Mn$_4$N/Pt (6.969 m J/m$^2$)[23]. The reason is that the calculated iDMI are based on Mn$_4$N [23] or Mn [33] ultrathin films while our measurement was performed on a 17nm thick film. The DMI we measured is the average DMI over the film. As mentioned previously, since the predicted iDMI is an interfacial effect and decays away from the surface, the average DMI decreases significantly as the thickness increases and much smaller than the predicted iDMI. To further investigate the relationship between our measured DMI and predicted DMI, we conducted a detailed comparison of the DMIs. In Figure.3(e), the red dots correspond to the average DMI ($D_{average}$) in MgO/Mn$_4$N/Pt and MgO/Mn$_4$N/Cu obtained from DFT calculations [23].

$$D_{average} = \int_0^{t_m} D(t)dt/t_m \tag{2}$$

$$D(t) = D_0 \begin{cases} e^{\frac{0.4-t}{0.4}}, & t > 0.4 \\ 1, & t < 0.4 \end{cases} \tag{3}$$

Where D(t) is the DMI distribution function which describes the exponentially decaying iDMI from the surface [44]. $D_0$ is the iDMI at the surface from DFT [23], as shown in Table (1), and t is the distance from the surface in nm. As seen in Figure.3(e), there is a clear discrepancy between the calculated DMI and the experimental DMI. One of the possible reasons is the presence of the mixing layers at the Mn$_4$N interfaces. In DFT, interface between two layers is assumed to be an ideal surface, which means that there are no atomic mixings. However, XPS and polarized neutron reflectometry (PNR) found that the interfaces of Mn$_4$N/Pt and MgO/Mn$_4$N are not ideal [41]. 3-4 nm mixing layers are present at the interfaces, including some MnO at the surface of the Mn$_4$N. These mixing layers at the interface decrease the accuracy of the iDMI from the DFT calculation, which can explain the discrepancy between the predicted DMI and the measured DMI.

**Table 1.** interfacial DMI at the surface of MgO/Mn$_4$N, Pt/Mn$_4$N (001) and Cu/Mn$_4$N (001), calculated from DFT[23]

| **MgO/Mn$_4$N (001)** | **Pt/Mn$_4$N (001)** | **Cu/Mn$_4$N (001)** |
|---|---|---|
| -1.017 mJ/m$^2$ | 6.969 mJ/m$^2$ | -2.633 mJ/m$^2$ |

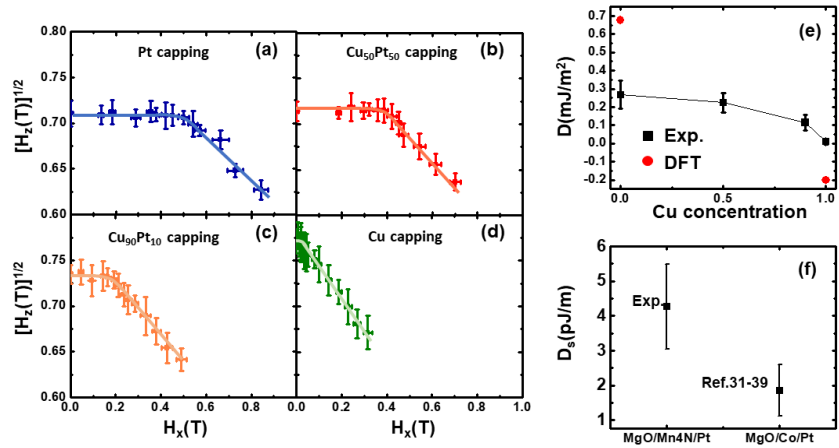

**Figure 3.** (a)-(d) $H_z^{1/2}$ vs. $H_x$ curves of MgO/Mn$_4$N/Cu$_x$Pt$_{1-x}$ with different capping layers x=0,0.5,0.9,1). The lines in the graphs are the fitted lines. The turning points, where $H_z^{1/2}$ begins to decrease linearly vs. $H_x$, correspond to the $H_{DMI}$ of each capping. (e) Measured DMI of MgO/Mn$_4$N/Cu$_x$Pt$_{1-x}$ with different Cu concentration in the capping layers. Red dots indicate the DMI from DFT calculations. (f) $D_s$ comparison between MgO/Mn$_4$N/Pt and MgO/Co/Pt [31-39].

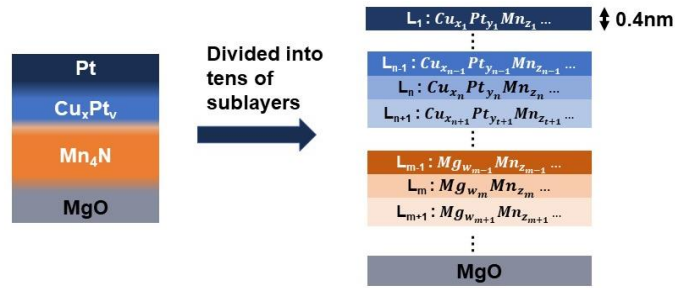

**Figure 4.** Schematic diagram of the solid solution model. A sample is divided into tens of sublayers with a thickness of 0.4 nm using compositional gradient.

*Mixing layer effect on DMI*

To investigate the possible effect of the mixing layers to the DMI, a solid solution model is built. In this model, the multilayer is divided into tens of sublayers with a thickness of 0.4 nm, as shown in Figure 4. The composition of each sublayer is estimated from XPS data using the method by Ma et al. [43]. There are two assumptions in this model. First, the DMI between two sublayers is originated from the interactions between (Cu, Pt, Mg) in one layer $(L_{n'})$ and Mn atoms in another layer $(L_n)$. This means that the DMI is proportional to the concentration of Pt$(y_{n'})$ in layer $L_{n'}$ and concentration of Mn$(z_n)$ of $L_n$. It is the same for the DMI between Cu and Mn, or Mg and Mn. Second, the DMI from one sublayer $L_n$ decays exponentially from the surface of layer $L_n$, which is given by equation (2).

The total DMI from Pt acting on a sublayer $L_n$ ($DMI_{Pt-L_n}$) is the sum of DMI of Pt from all over other sublayers:

$$DMI_{Pt-L_n} = \sum_{n',n'\neq n} z_n * y_{n'} * D_{n,n'} S_{n,n'} \tag{4}$$

$$D_{n,n'} = D_{0-Pt/Mn4N} \begin{cases} e^{\frac{0.4-|t_n-t_{n'}|}{0.4}}, |t_n - t_{n'}| > 0.4 \\ 1, |t_n - t_{n'}| < 0.4 \end{cases} \tag{5}$$

$$S_{n,n'} = \begin{cases} 1, t_n < t_{n'} \\ -1, t_{n'} < t_n \end{cases} \tag{6}$$

Where $D_{0-Pt/Mn4N}$ is the calculated DMI at Pt/Mn$_4$N interface [24], as shown in Table 1, and $S_{n,n'}$ is a function that assigns the direction of DMI. The DMI from the Pt atoms under the layer $L_n$ ($t_{n'} < t_n$) has an opposite sign comparing to the DMI from the Pt atoms above the layer $L_n$ ($t_n < t_{n'}$).

The total average DMI from Pt (DMI$_{Pt}$) is the sum from all layers n:

$$DMI_{Pt} = \sum_n DMI_{Pt-L_n} * 0.4/T \tag{7}$$

Where T is the total thickness of the samples. The same method was used to calculate the total average DMI from Cu (DMI$_{Cu}$) and from Mg (DMI$_{Mg}$). And the total DMI ($DMI_{tot}$) in the film is given in equation (7) by the sum of DMI from Pt, Cu and Mg.

$$DMI_{tot} = DMI_{Pt} + DMI_{Cu} + DMI_{Mg} \tag{8}$$

Figure.5(a)-(d) shows the compositional depth profiles of Mn$_4$N with various Cu$_x$Pt$_{1-x}$ capping layers obtained from analysis of XPS measurements. Figure.5(a) is obtained from our previous publication on the mixing layers in MgO/Mn$_4$N/Pt [41]. Here, the surface of the Pt

layer is set at z = 0 nm. The depth profile shows that elemental mixings exist at all the interfaces (Pt/$Cu_xPt_{1-x}$, $Cu_xPt_{1-x}$/$Mn_4N$, and $Mn_4N$/MgO). The diffusion from the protective Pt layer on top of the $Cu_xPt_{1-x}$ capping layer increases the actual Pt concentration in $Cu_xPt_{1-x}$, and consequently increases the Pt diffusion into $Mn_4N$. Based on the compositional depth profile in Figure.5(b)(c)(d), using z = 4 nm as a reference point, the actual Cu:Pt ratio of the capping layer is 41:54 for $Cu_{50}Pt_{50}$ capping sample. It is 64:22 for $Cu_{90}Pt_{10}$ capping sample and 77:13 for Cu capping sample. From all four profiles, oxygen peaks are found to exist near the top surface of $Mn_4N$ (z < 7nm) and the Mg:O ratio near the bottom surface (z > 20nm) is smaller than 50:50. These indicate that both $Cu_xPt_{1-x}$/$Mn_4N$ and $Mn_4N$/MgO interface have some MnO. The presence of MnO at the $Cu_xPt_{1-x}$/$Mn_4N$ originated from oxidization during the cooling process between the $Mn_4N$ deposition and CuPt deposition, and the MnO at the surface of $Mn_4N$/MgO is due to the oxygen diffusion from the MgO substrate to the $Mn_4N$ layer[41]. Using the compositional depth profile in Figure.5(a)-(d), the average DMIs of MgO/$Mn_4N$/$Cu_xPt_{1-x}$ were estimated by the solid solution model. Figure.5(e) shows the comparison between the calculated DMI using the solid solution model and the measured DMI. The calculated DMIs, which are indicated by the green triangles, are in agreement with the measured DMI, indicated by black squares.

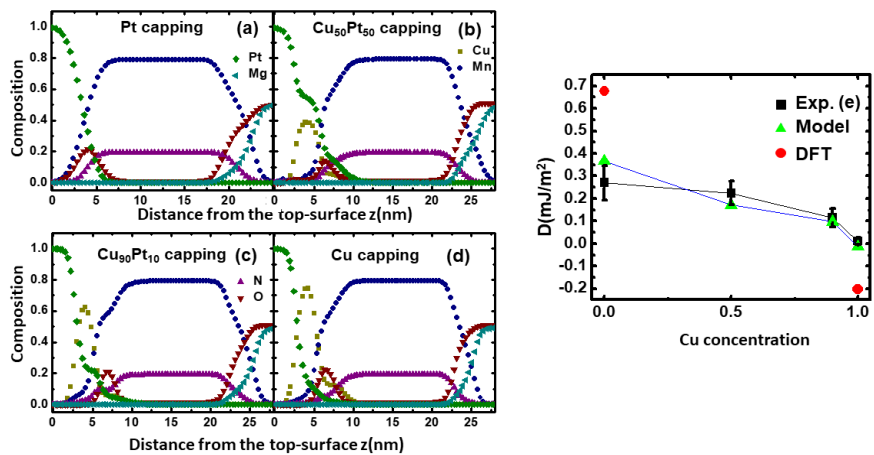

**Figure 5.** (a)-(d) Compositional depth profile of Mn$_4$N samples with different capping layers as a function of distance from the Pt protective layer obtained from analysis of XPS measurements. (a) is obtained from ref.41. (e) Comparison of calculated DMI from the solid solution model(green), measured DMI (black), and average DMI based on DFT (red).

## Conclusions

The Dzyaloshinskii-Moriya Interactions (DMIs) of MgO/Mn$_4$N/Cu$_x$Pt$_{1-x}$ multilayers were measured by extracting H$_{DMI}$ from the angular dependence of the coercivity field based on the magnetic droplet nucleation model. The compositional dependence of the DMI is non-linear in Cu concentration. The interfacial DMI constant D$_s$ of MgO/Mn$_4$N/Pt is larger than that of MgO/Co/Pt, which is consistent with the chemical trend of DMI among the transition metals. To study the effect of mixing layers on DMI, a simple solid solution model with mixing layers effect is built, based on the X-ray photoelectron spectroscopy (XPS) measurement, and the average DMI from this model is in good agreement with the measured DMI. Our experimental results provide a promising approach to control the DMI in Mn$_4$N-based thin films, with implications in achieving small skyrmion and enabling future spintronics technologies, and a method to connect the density functional theory (DFT) calculated DMI and measured DMI.

**Author Contributions:**

Wei Zhou: Sample fabrication, measurements, modeling, Writing – original draft

Chung Ting Ma: Modeling, Writing – review & editing

S. Joseph Poon: Funding acquisition (lead); Supervision (lead); Writing – review & editing.

**Funding:**

This work was supported by the DARPA Topological Excitations in Electronics (TEE) program (grant D18AP00009). The content of the information does not necessarily reflect the position or the policy of the Government, and no official endorsement should be inferred. Approved for public release; distribution is unlimited.

**Data Availability Statement:**

The data that support the findings of this study are available from the corresponding author upon reasonable request.

**Acknowledgments:**

The Phi VersaProbe III XPS used for acquiring the data was provided through the NSF-MRI Award #1626201.

**Conflicts of Interest:** The authors declare no conflict of interest.